\documentclass[a4paper,11pt]{article}

\usepackage{e-jc}

\usepackage{picture}
\usepackage{amsmath}
\usepackage{amssymb}
\usepackage{graphicx}
\usepackage{epstopdf}
\usepackage{listings}

\newcommand{\qed}{\nobreak \ifvmode \relax \else
      \ifdim\lastskip<1.5em \hskip-\lastskip
      \hskip1.5em plus0em minus0.5em \fi \nobreak
      \vrule height0.75em width0.5em depth0.25em\fi}

\author{Hyokun Yun\\
\texttt{yun3@purdue.edu}\\
Department of Statistics\\
Purdue University
}
\title{Using Logistic Regression to Analyze the Balance of a Game:\\
  The Case of StarCraft II\texttrademark}
\begin{document}

\lstset{language=R, breaklines=true}

\bibliographystyle{plain}
\maketitle

\section{Introduction}

\subsection{Motivation}

Recently,
the market size of online game has been increasing astonishingly
fast, and so does the importance of good game design.
In online games, usually a human user competes with others,
so the \emph{fairness} of the game system to all users is of great
importance not to lose interests of users on the game.
Furthermore, the emergence and success
of electronic sports (e-sports) and professional gaming
which specially talented gamers compete with others
draws more attention on whether they are competing in the
fair environment.

No matter how fierce the debates are in the game-design community,
it is rarely the case that one employs statistical analysis to
answer this question seriously. But considering the fact that
we can easily gather large amount of user behavior data on games,
it seems potentially beneficial to make use of this data to aid making
decisions on design problems of games.
Actually, modern games do not aim to perfectly design the game at
once: rather, they first release the game, and then monitor users'
behavior to better balance the game. In such a scenario,
statistical analysis can be particularly helpful.

Specifically, we chose to analyze the balance of StarCraft II\texttrademark,
which is a very successful recently-released real-time strategy (RTS)
game. It is a central icon in current e-Sports and professional gaming
community:
from April 1st to 15th, there were 18 tournaments of StarCraft II\texttrademark.
However, there is endless debate on whether the winner of the
tournament is actually superior to others, or it is largely due to certain
design flaws of the game. In this paper, we aim to answer such a
question using traditional statistical tool, logistic regression.

\subsection{Problem Setting}

In 1 vs. 1 match of this game, each gamer chooses
his/her race of army to play. There are three races: Terran, Protoss,
and Zerg. Note that it is allowed for two gamers to choose the same race.
Also, the map\footnote{Actually, it is more accurate to call it the
battleground of the war, but it is conventionally called as a
\emph{map}.}
is chosen, usually according to the rule of the tournament.
When races of two players and the map are chosen,
two gamers begin a war until
one gamer gives up and admits that he lost a game,
or certain end-of-the-game conditions are met.

In traditional games like chess or go,
two gamers are in perfectly same condition except
the right of the first move. In games like StarCraft II,
the gamer can choose very important characteristics of his/her army to
play,
so whether it is a fair game is an important issue in the community.
Particularly, people are extremely interested in whether
playing a certain race is particularly advantageous against the other.
For example, a lot of people argue that it is difficult for a
Zerg player to defeat a Terran player.

However, note that the balance between two races also depends heavily on
the map they are playing in. For example, there are maps which
bases of two players are located closely. It is generally conceived
that such a location of bases favors the Terran race,
because the Terran army is powerful but immobile relative to others.
But there are numerous other factors of the map design that designers of the map
can utilize to make the game balanced, and we are usually interested in
\emph{overall} effect of such factors.

\subsection{Data Description}

In this research, we used the result of 852 games in
Global Starcraft League\texttrademark (GSL)\footnote{
\texttt{http://wiki.teamliquid.net/starcraft2/GOMTV\_Global\_Starcraft\_II\_League}
}
from October 2010 to March 2011.
In this the most prestigious league of StarCraft II\texttrademark,
64 number of professional gamers
compete to each other, and the winner of the league
gets \$87,000.

Each record of data consists of the identifiers of two players who played
the game against each other, their corresponding choice of races,
the map which the game was played, the date of the match,
and the duration of the match: see Table~\ref{variable_table}.
Players rarely change his race between games,
since it requires a lot of pratice to be good at
just a single race.
Thus, one may view two variables as two levels of a hierarchical
information:
race provides higher-level information and player provides
lower-level information.
There was only one gamer who played randomly chosen race
for just two games until being eliminated from the tournament,
and his games were omitted from data.

There are 136 users and 14 maps in this data.
The data was gathered from the official website of
GSL\footnote{\texttt{http://esports.gomtv.com/gsl/}},
using Python-based web crawler we created on our own.

\begin{table}[h]
  \centering
  {\small
  \begin{tabular}{llll}
    \hline
    Variable & Type & Description & Example \\
    \hline
    Winner & Binary (0 or 1) & 1 if Player 1 won the game, 0 otherwise
    & 0, 1 \\
    Player 1 & Nominal & Identity of one player of the
    game & Jonathan Walsh  \\
    Race 1 & Nominal & The race of the Player 1 & Protoss \\
    Player 2 & Nominal & Identity of the other player of
    the
    game & Greg Fields \\
    Race 2 & Nominal & The race of the Player 2 & Zerg \\
    Map & Nominal & The map the game was played & Xel'Naga Caverns\\
    Date & Interval & The date of the match & Jan. 01, 2011 \\
    Duration & Interval & The duration of the match & 21 min 35 sec \\
    \hline
  \end{tabular}
  }
  \caption{List of Variables in the Data}
  \label{variable_table}
\end{table}

\section{Methods}

\subsection{Specification of Model}

Recall from Table~\ref{variable_table} that
$Winner_i = 1$ if Player 1 wins the game.
We model each $Winner_i$ to be independent Bernoulli random variable
with $\pi_i := P(Winner_i = 1)$ as:
\begin{equation}
  logit(\pi_i)
  :=
  \log \frac{\pi_i}{1-\pi_i}
  =
  \beta_{Player1_i} - \beta_{Player2_i} + \beta_{(Map, Race1, Race2)_i}
  .
  \label{model_eq}
\end{equation}

\begin{figure}[hbp!]
  \centering
  \begin{picture}(220,140)
    \put(10,0){\line(0,1){140}}
    \put(10,100){\framebox(209,30)[c]{$\beta_{Greg Fields} = 2.098$}}
    \put(50,60){\framebox(169,30)[c]{$-\beta_{Jonathan Walsh} = -1.797$}}
    \put(50,20){\framebox(7,30)[r]}
    \put(57,20){\makebox(100,30)[l]{ $\beta_{\text{Xel'Naga
            Caverns},Zerg,Terran} = 0.0632$}}
    \put(0,0){\makebox(100,30)[l]{0}}
    \put(10,15){\vector(1,0){47}}
    \put(57,20){\line(0,-1){10}}
    \put(57,0){\makebox(100,30)[l]{ $0.36336 \approx 2.098 - 1.797 +
        0.0632$}}
    \put(219, 100){\line(0,-1){40}}
    \put(50, 60){\line(0,-1){10}}
  \end{picture}
  \caption{Illustration of the Model}
  \label{howto_figure}
\end{figure}
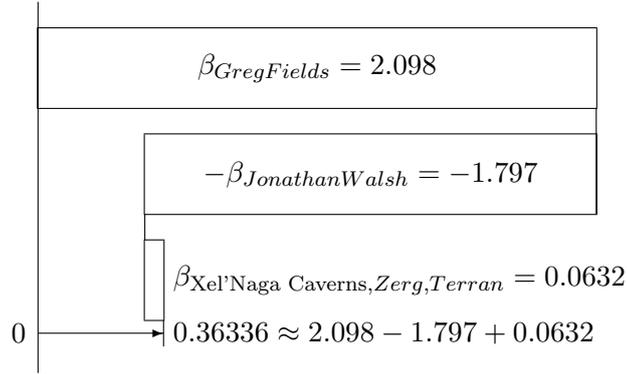

Let us try to undertand the intuition behind the model,
using an example with a figure.
Suppose that famous gamers Greg Fields and Jonathan Walsh are
playing in the map, Xel'Naga Caverns. Please refer to
equation \eqref{model_eq} and Figure~\ref{howto_figure}.

If player 1 is a great gamer, than $\beta_{Player1}$ will be high,
and it will increase $\pi$, consequently increasing the probability
that he will win the game. Greg Fields' is one of the greatest
Zerg player in the world, so his estimated $\beta$ is very high:
$\hat{\beta}_{Greg Fields} = 2.098$.

On the other hand, no matter how good the player 1 is,
if his opponent, player 2, is also a strong one, then the probability
he will win the game should certainly decrease.
It is being considred by $\beta_{Player2}$.
The opponent Jonathan Walsh is also one of the greatest
Terran players, and his estimated parameter
$\hat{\beta}_{Jonathan Walsh} = 1.797$. This value is subtracted from
$\beta_{Greg Fields}$, decreasing Greg's probability of
winning the game.

However, we should consider the map two are
playing in. It is Xel'Naga Caverns, and it turns out that
the map favors Zerg slightly over Terran. Thus,
$\beta_{(\text{Xel' Naga Caverns},Zerg,Terran)} = 0.0632$,
increasing the probability Greg wins on this map.

Finally, since $logit(\hat{\pi}_i) = 0.36336$, we may transform the
logit function to get
\begin{equation}
  \hat{P}(\text{Greg Wins}) := \hat{\pi}_i = \frac{1}{1+\exp(-0.36336)} = 0.516,
\end{equation}
using standard mathematical operation used frequently in logistic regression.
Since both are very good players,
it turns out that it is very hard to predict who will win the game.
This is sensible, since the game is by its nature not very
predictable. If we could easily predict the result of the game,
nobody would want to watch the game to waste time.
Using statistical analysis, however, we can get the
overall tendency in games, even though the signal may not be
very strong. In this case, the fitted model tells you it is
more likely for Greg to win over Jonathan.

Note that we are assuming each games to be \emph{conditionally}
independent, given players and map of the game.
The possible problems of this assumption will be discussed
with other limits of the model in Section~\ref{limit_section}.

\subsection{Restrictions on Parameter Space}

Note that when the position of Player 1 and Player 2
is switched in equation \eqref{model_eq},
then it should still give us an equivalent result.
In the former example of Greg and Jonathan,
as $\hat{P}(\text{Greg Wins}) = 0.516$, $\hat{P}(\text{Jonathan Wins})$ should be
$1 - 0.516 = 0.484$. To impose such a restriction, we need following
conditions: for each map, we should have
\begin{align}
  \beta_{Map, Terran, Zerg} &= -\beta_{Map, Zerg, Terran}, \\
  \beta_{Map, Terran, Protoss} &= -\beta_{Map, Protoss, Terran}, \\
  \beta_{Map, Protoss, Zerg} &= -\beta_{Map, Protoss, Zerg}.
\end{align}
Such a restriction can be naturally done within the framework of
standard logistic regression, not making the inference step
any harder. One has to use carefully designed data matrix,
but details are omitted since it is pretty straightforward.

\subsection{Characteristics of the Model}

\subsubsection{Advantages over Traditional Approach}

Let us discuss why we need such a statistical model
to analyze this data.
To compare the performance of each players,
the traditional way of analyzing the result of games is
to calculate the win rate of individual players.
For example, one may calculate
\emph{the fraction of game Greg won} and
\emph{the fraction of game Jonathan won},
and compare two numbers.
But this method is obviously problematic,
since in such a calculation, it does not distinguish
how strong one's opponent had been.
If one gamer has only encountered weak gamers by chance,
and have not yet been challenged by strong ones,
then would you still think he is a good gamer only because
he has a good win rate? Certainly not.
The beauty of having a statistical model is that we can
take care of this.
For example, Greg won only 41.67\% of the game,
while Jonathan won 50\%. But the fitted model does not tell
Jonathan is a better player, due to their respective history
of match: instead, it tells Greg is generally a better player,
with its parameter values
$\hat{\beta}_{Greg Fields} = 2.098$ and
$\hat{\beta}_{Greg Fields} = 1.797$.

On the other hand, it is always a hot debate whether
a certain map is balanced or not. But this is a hard question to
answer,
since you cannot simply say that the map named Xel'Naga Caverns
favors Zergs, since
many Zergs are winning over Terrans in this map.
Maybe we have not seen good enough Terrans playing in this map.
Or, we have not seen enough observations in this map.
When the \emph{fraction} of games Zerg has won over Terran is
calculated (for example, = 0.6), what is its standard error?
It is hard to answer, since each game is not
\emph{marginally} independent of each
other. We know that good gamers are more likely to win,
while bad gamers are less likely to do so.
However, it is much more reasonable to assume that they are
\emph{conditionally} independent, which is our assumption,
and in this case we can estimate the variability of our estimates.

Finally, it is hard to \emph{combine} estimates in traditional
approach. When Jonathan is playing a game with Greg in
Xel'Naga Caverns, how would you combine both gamers'
win rate and the fraction of Zerg won over Terran in Xel'Naga Cavern?
Usually, people stop to be \emph{quantitative},
and follow the \emph{qualitative} approach. In our model,
we can quantitatively combine fs to calculate overall effect.

\subsubsection{Limits}
\label{limit_section}

No matter how more attractive the model is compared to traditional
approaches, it is by no means a perfect model. To list why:

\begin{itemize}
\item \emph{Constancy of Parameter Values over Time}:
  The strength of each gamer is not constant over time.
  As a gamer accumulates experience, one generally gets
  better and better. On the other hand, it has been frequently
  observed that once legendary gamers become plain ones
  as they grow old and they cannot react as quickly as
  younger gamers. Thus using a single $\beta$ parameter
  for every game is actually problematic.
  However, in this data such an assumption was inevitable
  since we have only observed for seven months.
  When a study with larger longitudinal scale is done,
  we may even attempt to model this time-series effect.

\item \emph{Conditional Independence}:
  Each game may not be even conditionally independent
  given both players and map. For example, when two gamers
  are playing a best-of-five match, then the result of
  the first match certainly affects the second.
  When a player loses the first match, the player may get
  depressed, or having already seen which kind of style
  his opponent played, he may adjust his style very well
  and win the following game.   But since most of the match was played as
 a league match or
  as a best-of-three match, we assume that such a
  dependence between games is not very strong.
  Also, since we are having a lot of games (852),
  such a dependence between three or five games
  may not play a very significant role.

\item \emph{Interaction Between Players}:
  Everyone also knows that there should be a certain
  interaction effect between players.
  For example, a player with aggresive style is
  hard to win over a very defensive gamer.
  However, a defensive gamer may have hard time
  fighting agains a gamer who exploits the fact
  his opponent is defensive, and make a expansion
  very quickly. Since we have only 852 games and
  there are 136 gamers in the data, it is not
  possible to estimate $136 \times 136 = 18496$
  parameters with 852 numbers of games.
  However, we may partially take this into account
  using mixed-membership or latent feature models.
  We leave this interesting possibilities 
  for future work.

\item \emph{Interaction between Player and Map}:
  Sometimes it is clearly seen that certain user is
  very good at certain kinds of map. But although this kind of
  interaction is more tractable to deal with compared to
  player-player interaction, consideration of such factors
  are left for future work.

\end{itemize}

\subsection{Model Parsimony}

Although we have been trying hard to keep our model simple,
we still have too many parameters, since we give each player
one parameter. Since some gamers played only one or two games
just to lose and then be eliminated from the tournament, modeling even
such gamers will result in over-fitting of the problem and
numerical instability. Such a problem can generally be
taken care of by using \emph{regularized estimation} approach,
but it is slightly out of the scope of the course and
we lose the notion of \emph{probability} there.

Instead, one may try to do the \emph{variable selection}
himself/herself, not relying on indirect regularized estimation.
In general data analysis problems,
this is hard to do since it is hard to consider
every combination of variables. However, it is not that difficult
when one has good idea of which variables are not very necessary,
and it turns out to be
the case here: we \emph{know} that the players with small number of
games are problematic. Thus, we fix $\beta$'s of such players
to be 0. That is, the model \emph{gives up} to estimate
the performance of gamers who have not yet played enough, since
we do not have enough data. However, those gamers are
\emph{collectively} taken into account, since they do affect
the estimation of performance of other players and
balance of the map. And it has very natural interpretation:
when two players with not enough data are playing against
each other, the reasonable prediction is that it is a
50-50 game. When we know what map they are playing in,
then we may use the overall trend in the map to predict
the result, not being able to use further information at all.
It turns out to give similar result compared to
the use of $L_1$ regularization (lasso), which will be discussed in
the next section\footnote{Note: the model is unidentifiable by itself, but by setting
parameters for some players to 0, it becomes identifiable.
It is identifiable without such a treatment in lasso case.}.

\section{Results}
\subsection{Adequacy of the Model}

Firstly, the lack of fit was tested:
as a first step, it was done against constant model ($\beta_j =
0$ for all $j$), and the $p$-value was $10^{-6}$, naturally rejecting
the
null hypothesis that the constant model suffices in almost any
significance level.
Since this is almost
always the case for the data with considerable size, we also conducted
Hosmer-Lemeshow test\footnote{\texttt{R} code in
\texttt{http://www.stat.purdue.edu/\textasciitilde ovitek/STAT526-Spring11\_files/4-logistic.R}
were used}, using 10 groups: the $p$-value was 0.153, again favoring
our model\footnote{In this case, null hypothesis is our model.}.

Secondly, more modern method of cross-validation was used to
evaluate the quality of the fit. We conducted 10-fold cross-validation
and evaluated accuracy of our predictor in both training and test
data. Note that in our case, losses of type I and II
\emph{prediction} error are the
same (symmetric loss), so it suffices to check accuracy, unlike general cases.
The average accuracy in training data was $0.727 \pm 0.00797$
(mean $\pm$ standard deviation),
while that of test data was $0.706 \pm 0.0632$.
It seems our model generalizes quite well
(just 2\% drop of accuracy), but standard deviation is a bit high.
The reason should be that we do not have enough data about
every player: for some players with small number of games,
training data may not contain enough information about them
and cause inaccuracy of estimation for them, although on average
the model works quite well. To see whether overfitting is problematic
here, we've also used $L_1$ penalty (lasso) to estimate parameters
and evaluated its accuracy on exactly same partitions of data.
The accuracy was $0.708 \pm 0.0676$, which is not significantly
different from the model using no penalty. We did not set any
parameter to be 0 for lasso, but the set of nonzero parameters
chosen by lasso using another 10-fold cross-validation on training set
was very similar to what we've done by using the
number of games each gamer played: lasso also removed 94\% of players
we removed. In conclusion, overfitting was not a big problem,
and our selection of variables was not as ad-hoc as it could have
sounded.

\begin{figure}[t]
  \centering
  \includegraphics[width=45mm]{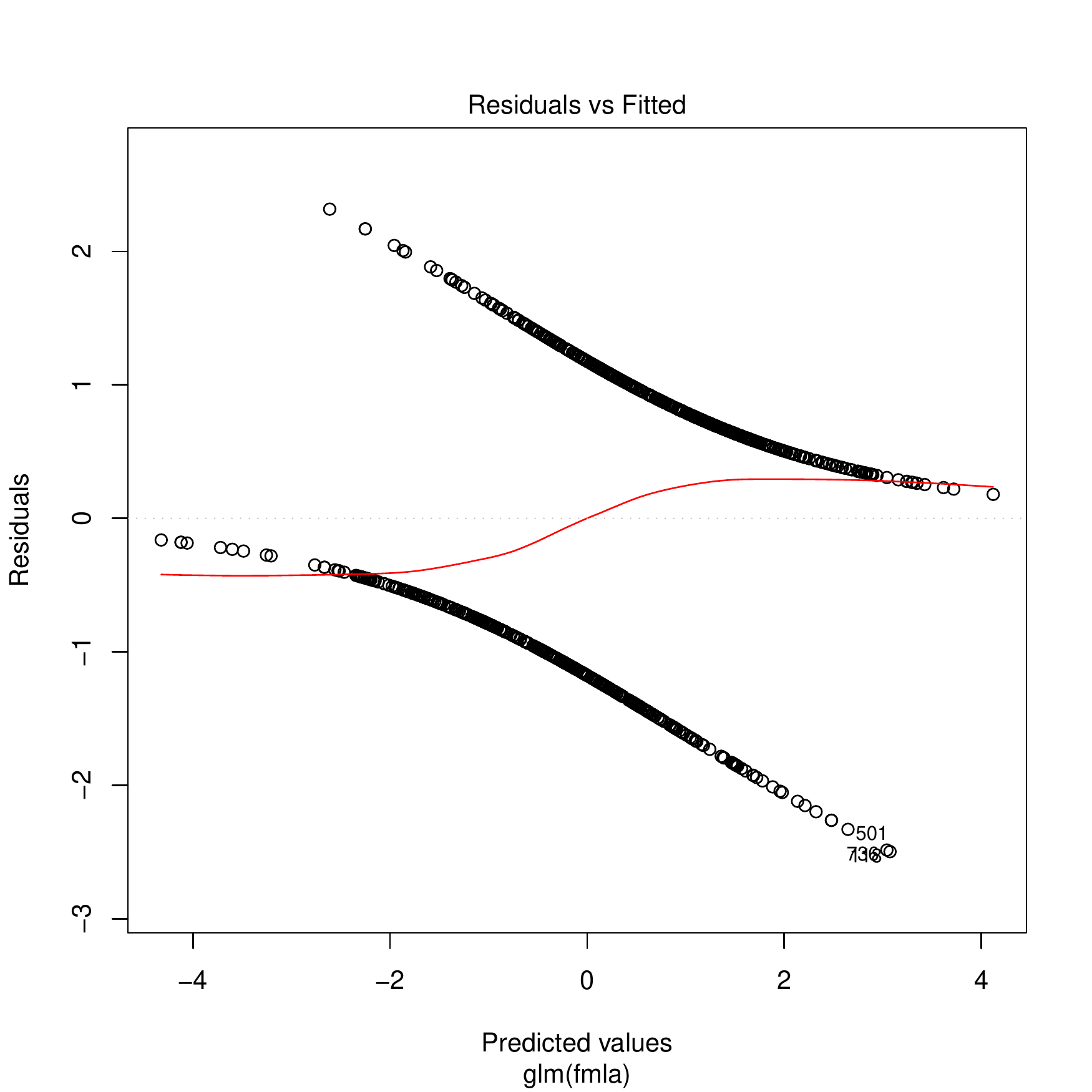}
  \caption{Residual vs. Fitted Plot}
  \label{residual_plot}
\end{figure}

Thirdly, we visually checked the quality of fit.
See Figure~\ref{residual_plot}. The plot is not as flat as
it is desired to be. This is because our fit is not perfect:
sometimes we predict a player to win by high probability,
which does not turn out to be the case.
Although we did not suffer overfitting in terms of accuracy,
the lack of regularizers may result in overfitting of
estimated probabilities, sometimes being overconfident
when the model should not be. Such an overfitting may
naturally occur when dealing with mid-sized data like this.
The use of regularization does not really help,
since in that case we lose the notion of probability.

Lastly, we checked whether our conditional indepdendence assumption
was adequate. When quasi-binomial model was fit, the dispersion
parameter was only 1.158738. To estimate the distribution of
estimated dispersion parameter, we bootstrap sampled 1000 datasets.
The mean and standard deviation of estimates were $1.275 \pm 0.284$,
clearly indicating that there does exist over-dispersion,
but the magnitued is not very serious.

\subsection{Interpretation}

\begin{figure}[t]
  \centering
  \includegraphics[width=40mm]{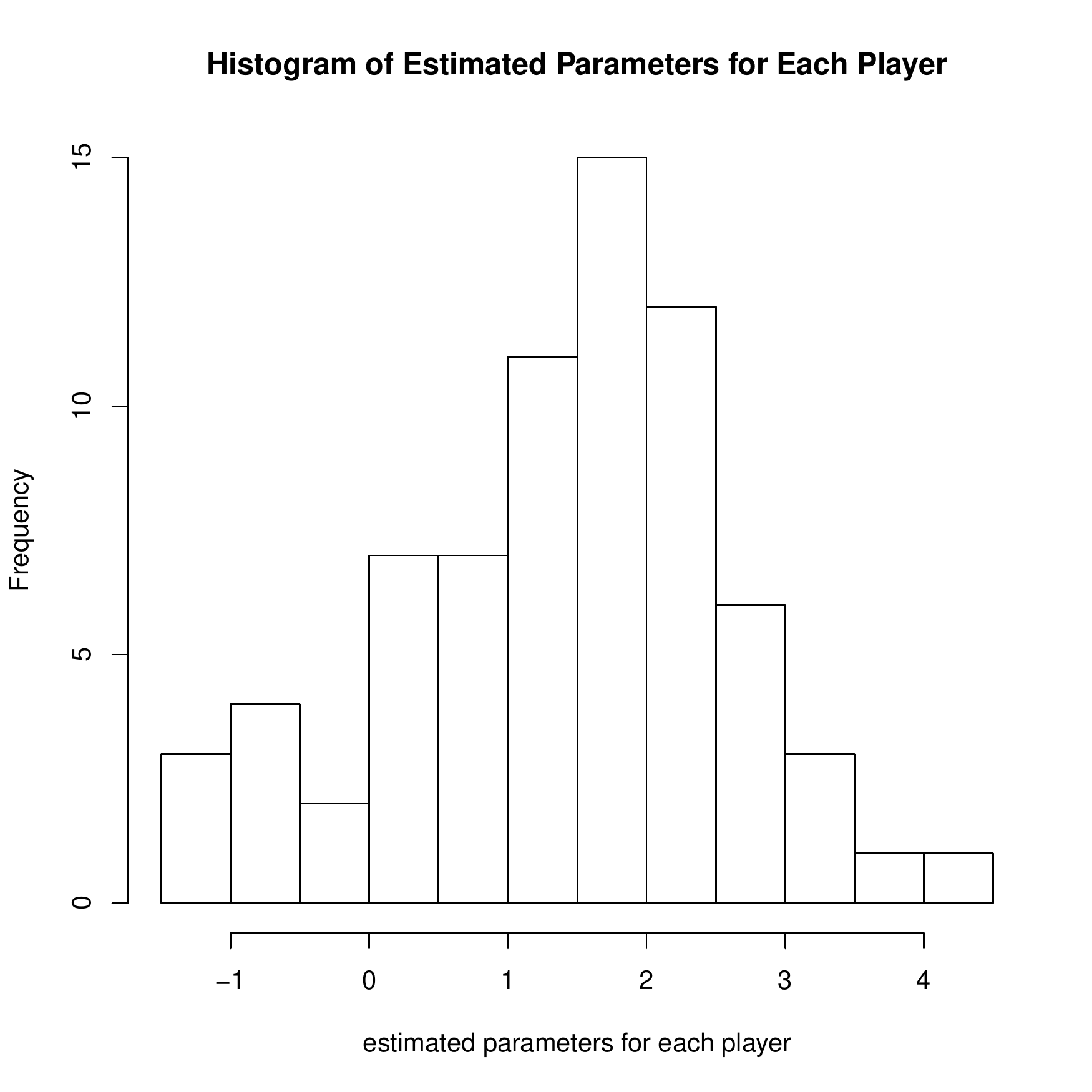}
  \includegraphics[width=35mm]{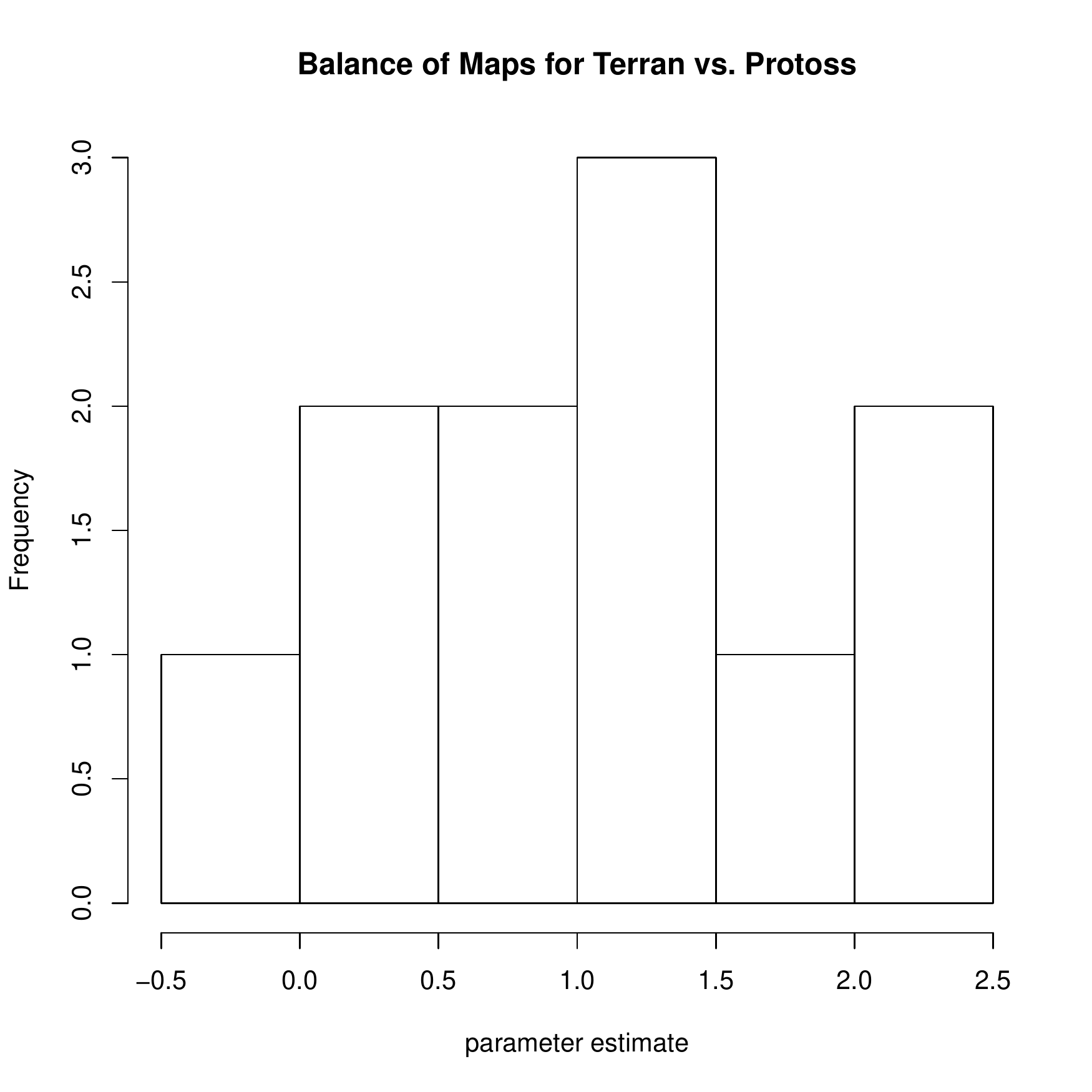}
  \includegraphics[width=35mm]{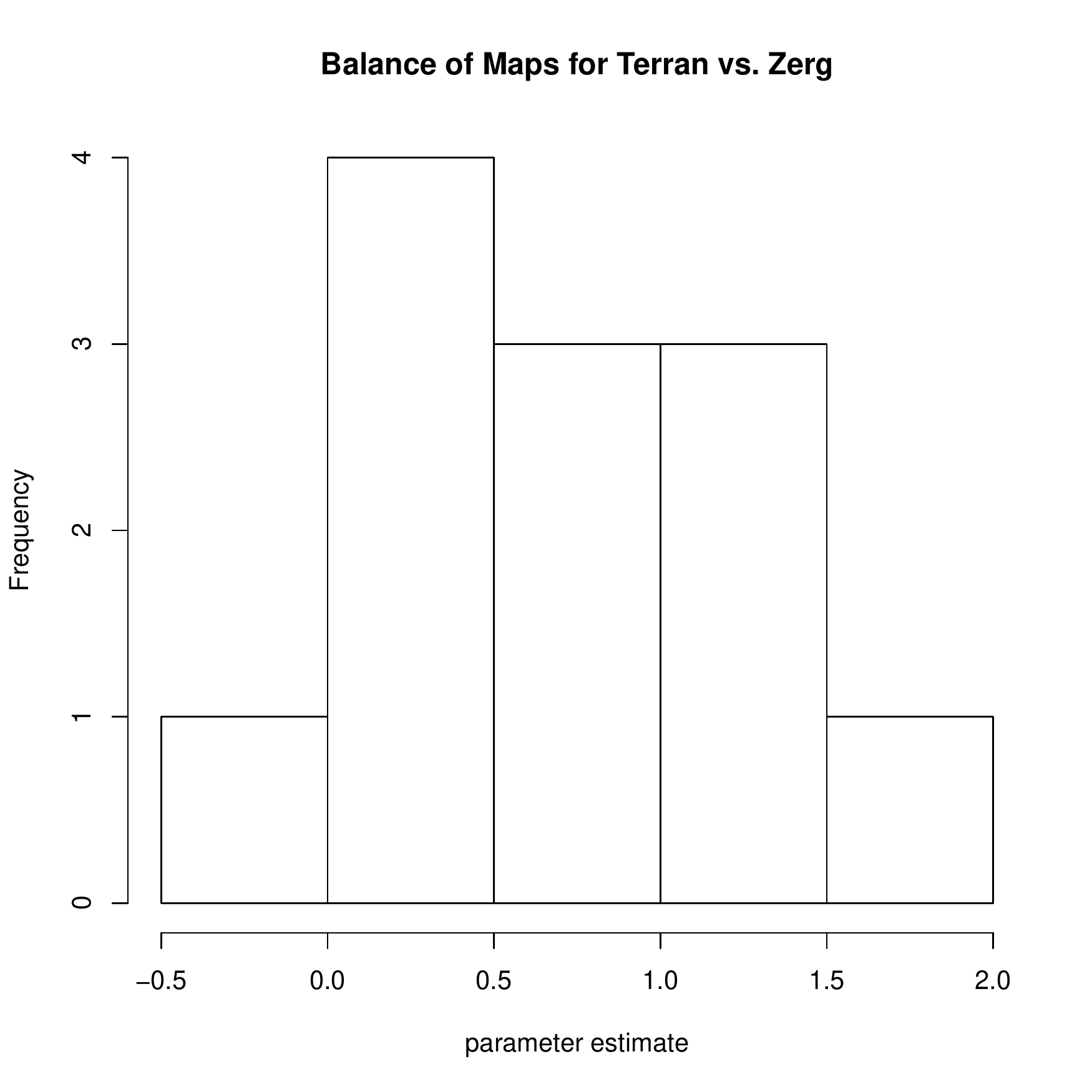}
  \includegraphics[width=35mm]{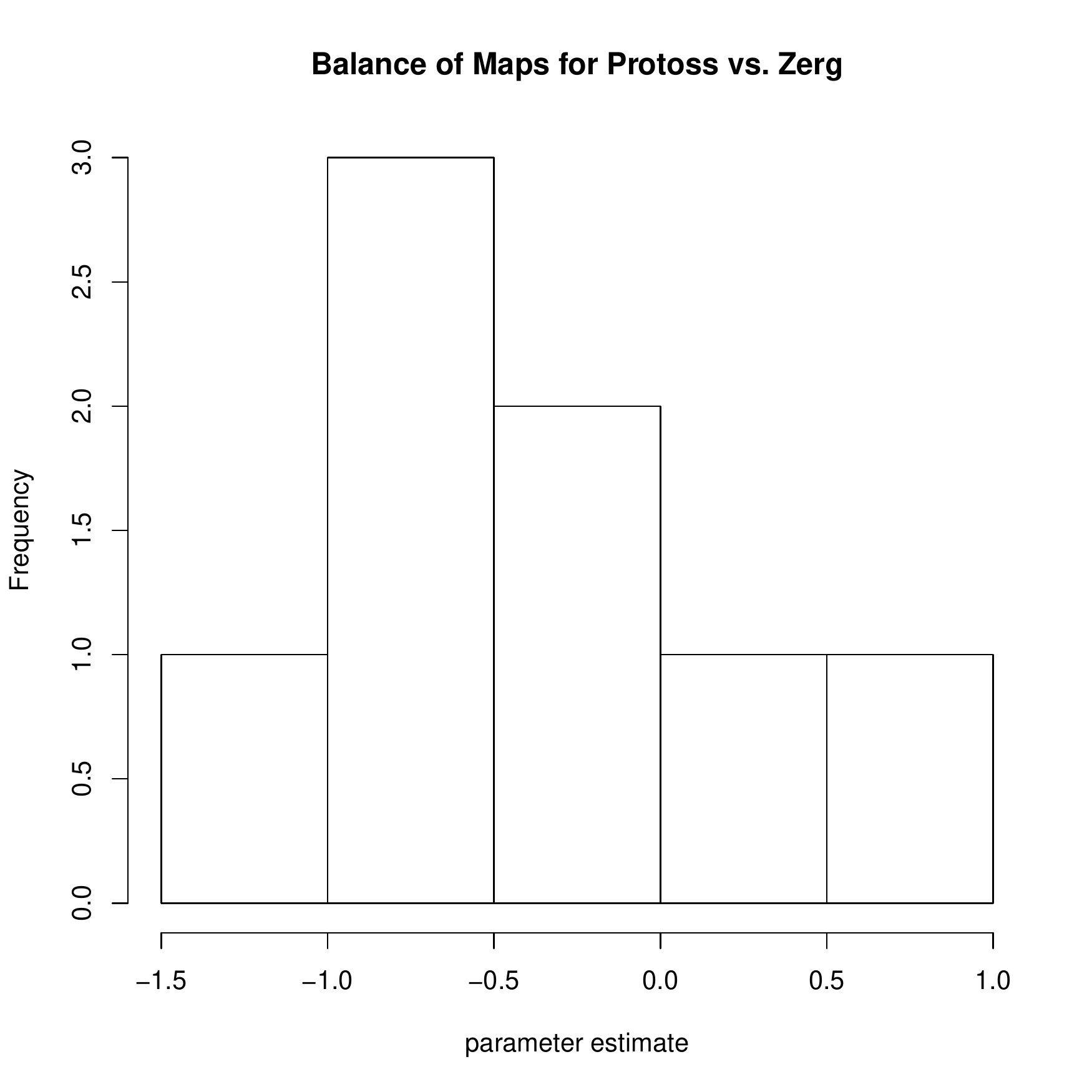}
  \caption{Estimated Parameters}
  {\small
  From left: (1) Estimated parameter for
  each player,
  (2) Estiamted parameters for Terran
  vs. Protoss
  in each map,
  (3) Same plot for Terran vs. Zerg.
  (4) Same plot for Protoss vs. Zerg.
  }
  \label{parameters_plot}
\end{figure}

Recall that there is one parameter given to each player,
which evaluates relative performance compared to others.
Figure~\ref{parameters_plot}. (1) plots estimated
parameter for each player: it is naturally centered in the point
bigger than zero, implying players with enough information are
better than those players whose parameters were set to be zero
because they did not play enough games. For interested readers
about ranks of parameters, refer to Table~\ref{ranks}.

In Figure~\ref{parameters_plot}. (2) to (4),
parameters which estimate the balance between two races
for each map were displayed. Since there are only 14 maps,
the histogram is very spiky. Mean and standard
deviation of estimates regarding Terran vs. Protoss,
Terran vs. Zerg, and Protoss vs. Zerg balance of map
was respectively
$1.064 \pm 0.821$, $0.749 \pm 0.566$, and $-0.369 \pm 0.596$.
It seems like the balance depends on the map,
but most maps favor Terran over Protoss and Zerg,
while the balance between Protoss vs. Zerg seemed
more adequate than others.

\begin{figure}[t]
  \centering
  \includegraphics[width=45mm]{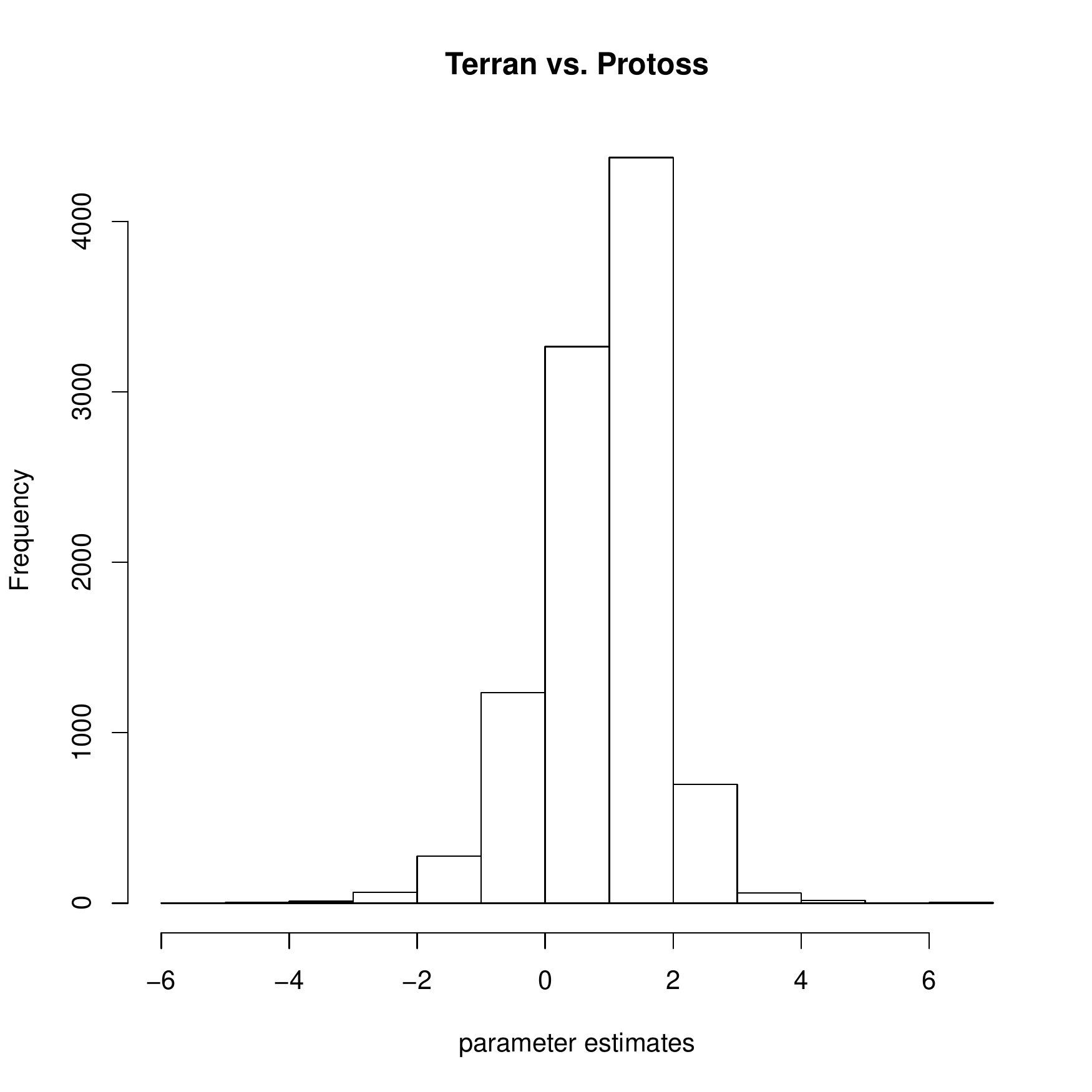}
  \includegraphics[width=45mm]{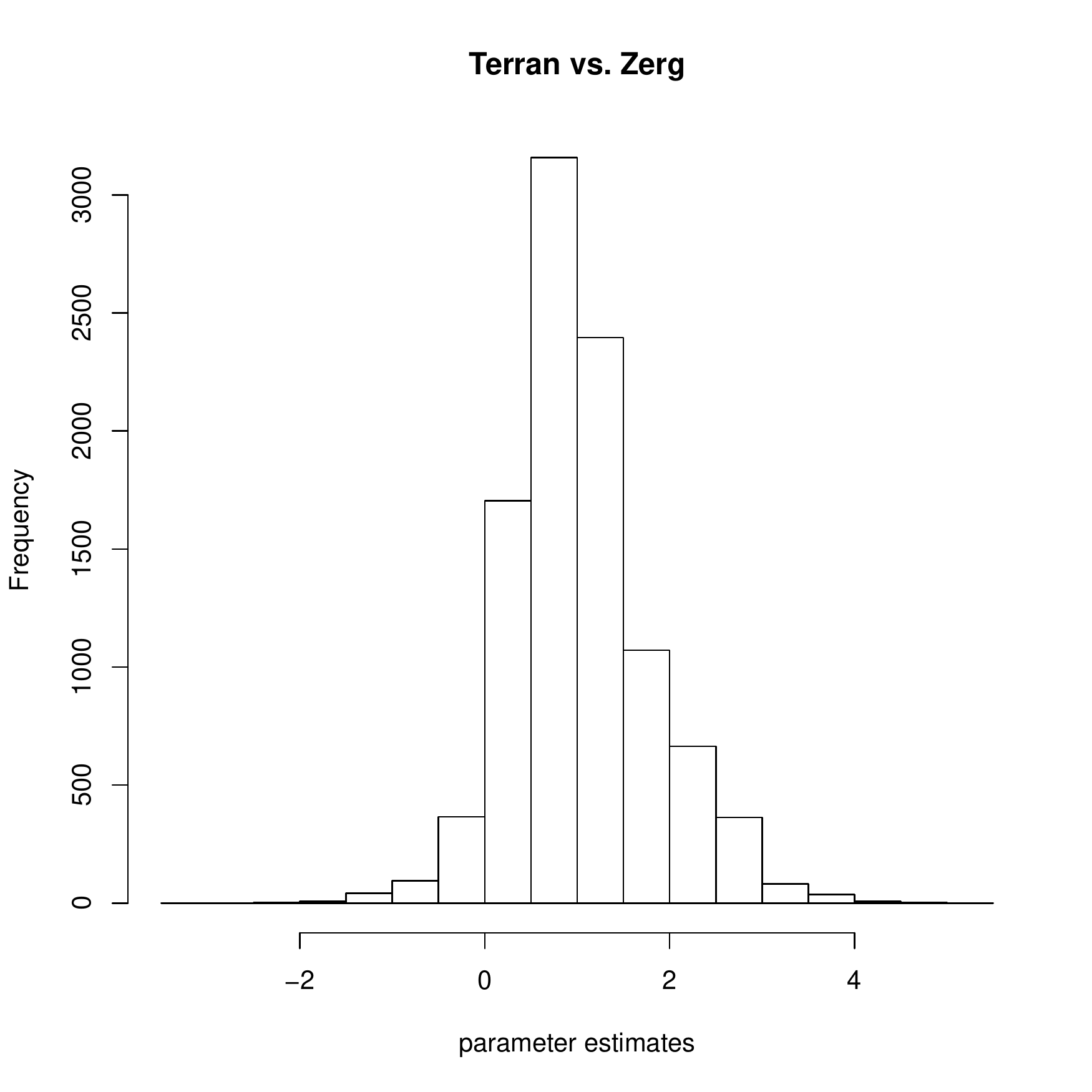}
  \includegraphics[width=45mm]{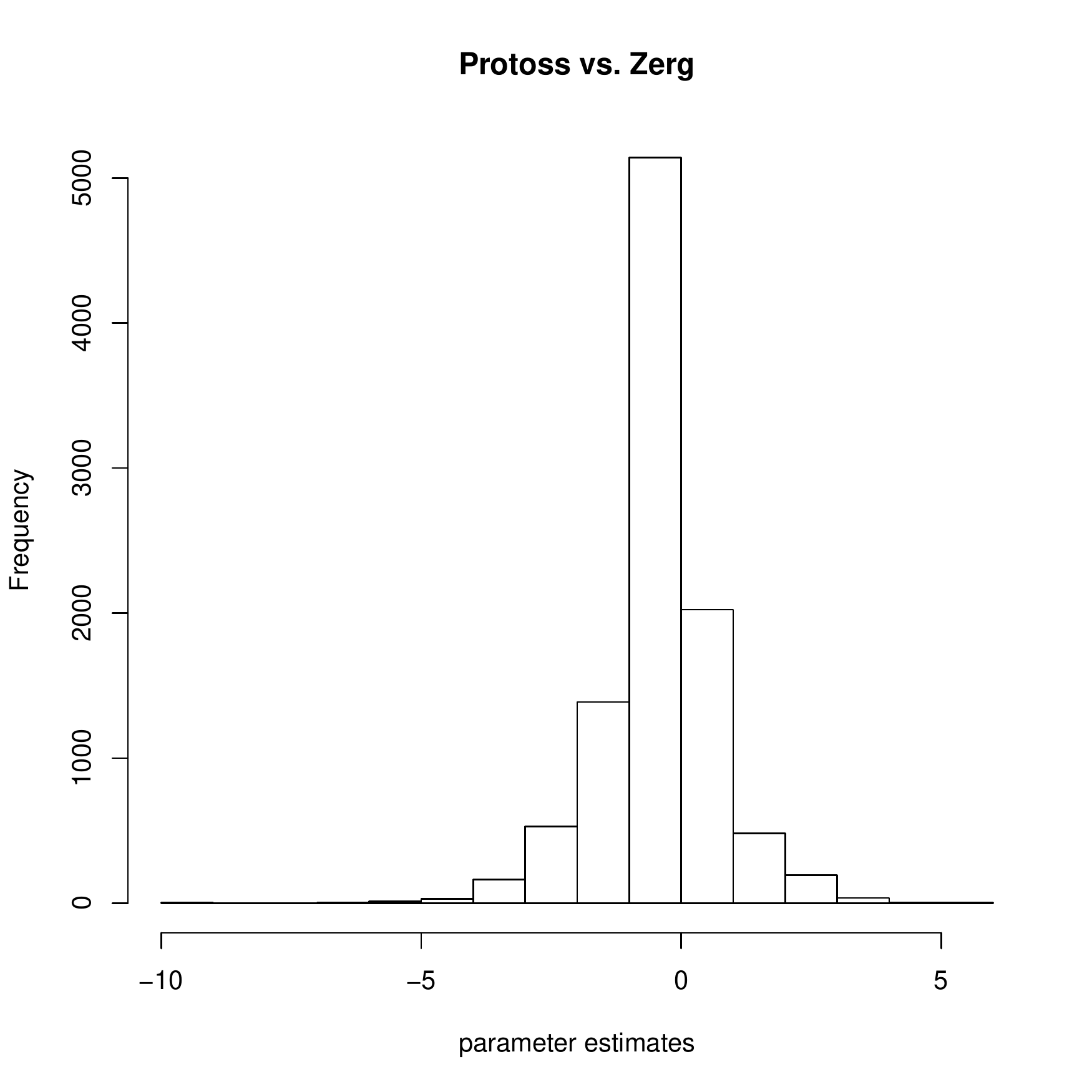}
  \caption{Bootstrap distributions of parameter estimates (Boostrap
    sample number: 10000)}
  \label{bootstrap_hist}
\end{figure}

To answer the higher-level question of
``So, is the game well-balanced?'',
we need to average over maps, since maps already
take balances into account individually.
Note that it is similar to testing
hypothesis about the overall mean in cell-means model of one-way ANOVA.
The parameter we test is:
\begin{equation}
  \beta_{Race1, Race2} = \frac 1 m \sum_{i=1}^m \beta_{Map_m, Race1, Race2},
  \label{mean_balance}
\end{equation}
for each $Race1, Race2$.
Since logistic regression does not have closed-form solutions of
parameter distributions, we bootstrap sampled 10000 datasets
and estimated \eqref{mean_balance} for each race combination.
As a result,
$P(\beta_{Terran, Protoss} > 0) \approx 0.839$,
$P(\beta_{Terran, Zerg} > 0) \approx 0.948$,
and
$P(\beta_{Protoss, Zerg} > 0) \approx 0.290$,
which implies that the unbalance between races are not very
significant in 5\% significance level even when each hypothesis
that the parameter value is exactly zero is tested individually
(when multiple hypotheses are simulatenously checked,
the significance level of the test drops).
However, certainly indications were seen that there may be some
balance problems, especially in the case of Terran vs. Zerg.
Interested readers may refer to histograms of estimates:
Figure~\ref{bootstrap_hist}.

\section{Discussion}

To authors' knowledge, this is the first time a standard statistical
technique which is more complex than mere summary statistics were used
to
analyze user behavior data in online games.
Using our technique, game designers may make use
of the results they gained from
beta-testers more carefully to reduce the cost of testing.
Especially in the past with StarCraft I, many times very
unbalanced maps were sometimes used in the tournament,
causing some strong players to be eliminated even early in the
tournament. Our model would be very helpful to prevent such a
disaster.

Since we've already discussed many of the technical problems
in above sections,
we conclude this section briefly.

\newpage

\section{Appendix}

\subsection{Appedix A : Descriptive Statistics}

\subsubsection{Number of races}
\begin{figure}[hbp!]
  \centering
	\includegraphics[width=55mm, height=55mm]{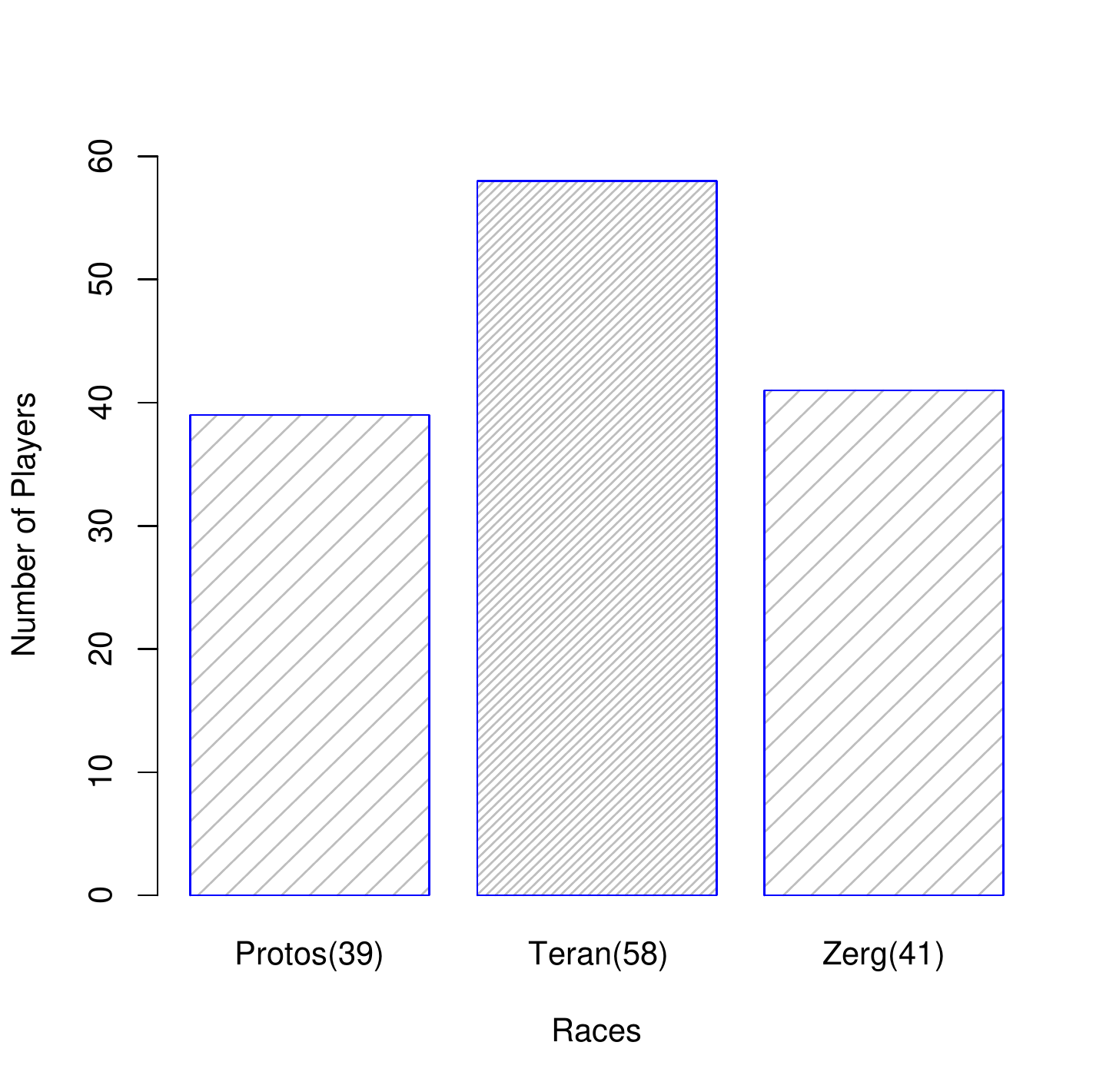} 
	\caption{Number of Races}
	\label{Number of Races}
\end{figure}

There are three types, \textit{Protos, Teran, Zerg}, of races in the Starcraft II. From the raw data, we found three observation whose race is \textit{r}. Only one player (ID : \texttt{GuMihofOu} ) used option \textit{random} (rare) which randomly assigns one of three races. Since he had played only three games in total, we eliminated these observations. After this elimination, we have \textbf{136 players}. The bar graph for race of these 136 players is shown in Figure~\ref{Number of Races}.

There is an outstanding preference to \textit{Teran} which possesses 42.6\% of the total players. At this point, balancing between races can be issued. We want to analyze this balancing problem using statistical approach.

\subsubsection{Number of observations (games) per player}
\begin{figure}[hbp!]
  \centering
	\includegraphics[width=150mm, height=90mm]{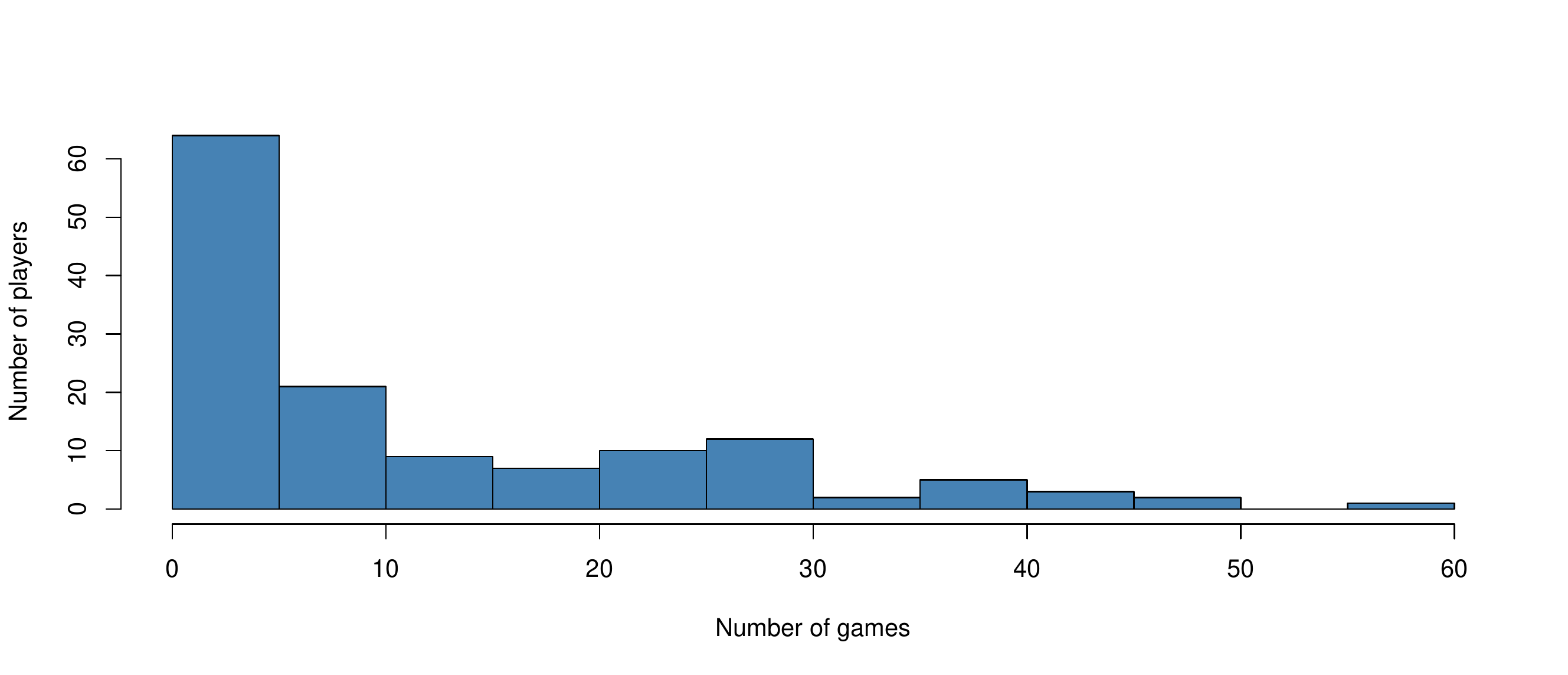}
	\caption{Game frequencies}
	\label{Game frequencies}
\end{figure}

There are 852 observations in the data set. The average number of games of each player is 6.26. However, it is well-known that a better player plays more games than others. In other words, there should be a large deviation of the number of games of the players. We observe this using a histogram in Figure~\ref{Game frequencies} and Table~\ref{Number of games of players}

\begin{table}[hbp!]
	\centering
	\begin{tabular}{c | r || c | r}
	\hline
	games &	glayers	& games	& glayers \\
	\hline
		1-5	&			64	& 31-35	& 2 			\\
 	6-10	& 		21	& 36-40	& 5				\\
	11-15	& 		9		& 41-45	& 3				\\
	16-20	& 		7		& 46-50	& 2 			\\
	21-25	& 		10	& 51-55	& 0 			\\
	26-30	& 		12	& 56-60	& 1 			\\
	\hline
	\end{tabular}
	\caption{Number of games of players}
	\label{Number of games of players}
\end{table}

\begin{figure}[hbp!]
  \centering
	\includegraphics[width=150mm, height=60mm]{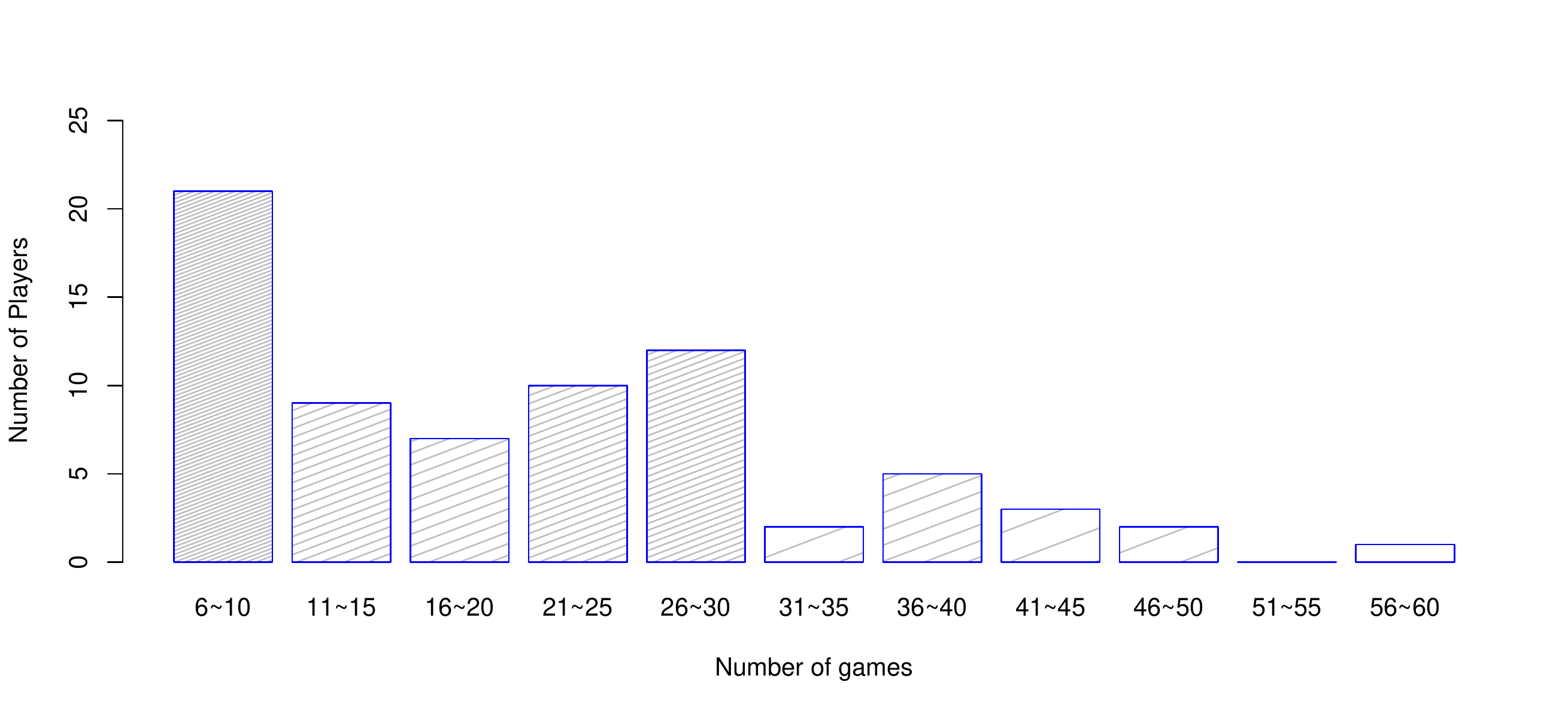}
	\caption{Number of players vs. Number of games}
	\label{playergames6}
\end{figure}

64 players (47\%) played only 1-5 games. Especially, 38 players (27.9\%) played only at most 2 games. Statistical results based on such players may not reliable. Hence we need to consider data reduction. For example, a bar graph of players who had played more than 5 games is in shown in Figure~\ref{playergames6}.

\subsubsection{Game frequencies between races}
Table~\ref{Number of games of race combination} shows how many games are done between each race combination. Notice that each combination is not in order. For example, the frequency of \texttt{Protoss vs. Teran} covers both \texttt{(player1,player2) = (Protoss, Teran)} and \texttt{(Teran, Protoss)}.

\begin{table}[hbp!]
	\centering
	\begin{tabular}{c | r }
	\hline
	Races	& Frequency \\
	\hline
	Protoss vs. Protoss &	45 \\
	Protoss vs. Teran	& 233 \\
	Protoss vs. Zerg &	131 \\
	Teran vs. Teran	& 134 \\
	Teran vs. Zerg	& 265 \\
	Zerg vs. Zerg	& 44 \\
	\hline
	\end{tabular}
	\caption{Number of games of race combination}
	\label{Number of games of race combination}
\end{table}

For balancing analysis purpose, we reduce our focus on battles between different races.

\subsubsection{Game frequencies between difference races}

Table~\ref{Game frequencies between difference races} shows the number of games of different races.

\begin{table}[hbp!]
	\centering
	\begin{tabular}{c | c | r | r }
	\hline
	Race vs. Race & Frequency &\multicolumn{2}{c}{Number of Wins} \\
	\hline
	Teran vs. Protoss &	233 & Teran: 121 & Protoss: 112 \\
	Teran vs. Zerg	& 265 &Teran: 132 & Zerg: 133 \\
	Protoss vs. Zerg &	131 & Protoss: 67 & Zerg: 64 \\
	\hline
	\end{tabular}
	\caption{Game frequencies between difference races}
	\label{Game frequencies between difference races}
\end{table}

As an intuitive check, we can compare the win/loss ratios,
$\frac{121}{121+112}=0.5193133, \frac{132}{132+1333}=0.4981132$ and 
$\frac{67}{67+64}=0.5114504$.
At a glance, those ratios do not look considerably apart from 0.5

\subsubsection{Time trend of the number of players for each race}
\begin{figure}[hbp!]
  \centering
  \includegraphics[width=130mm, height=90mm]{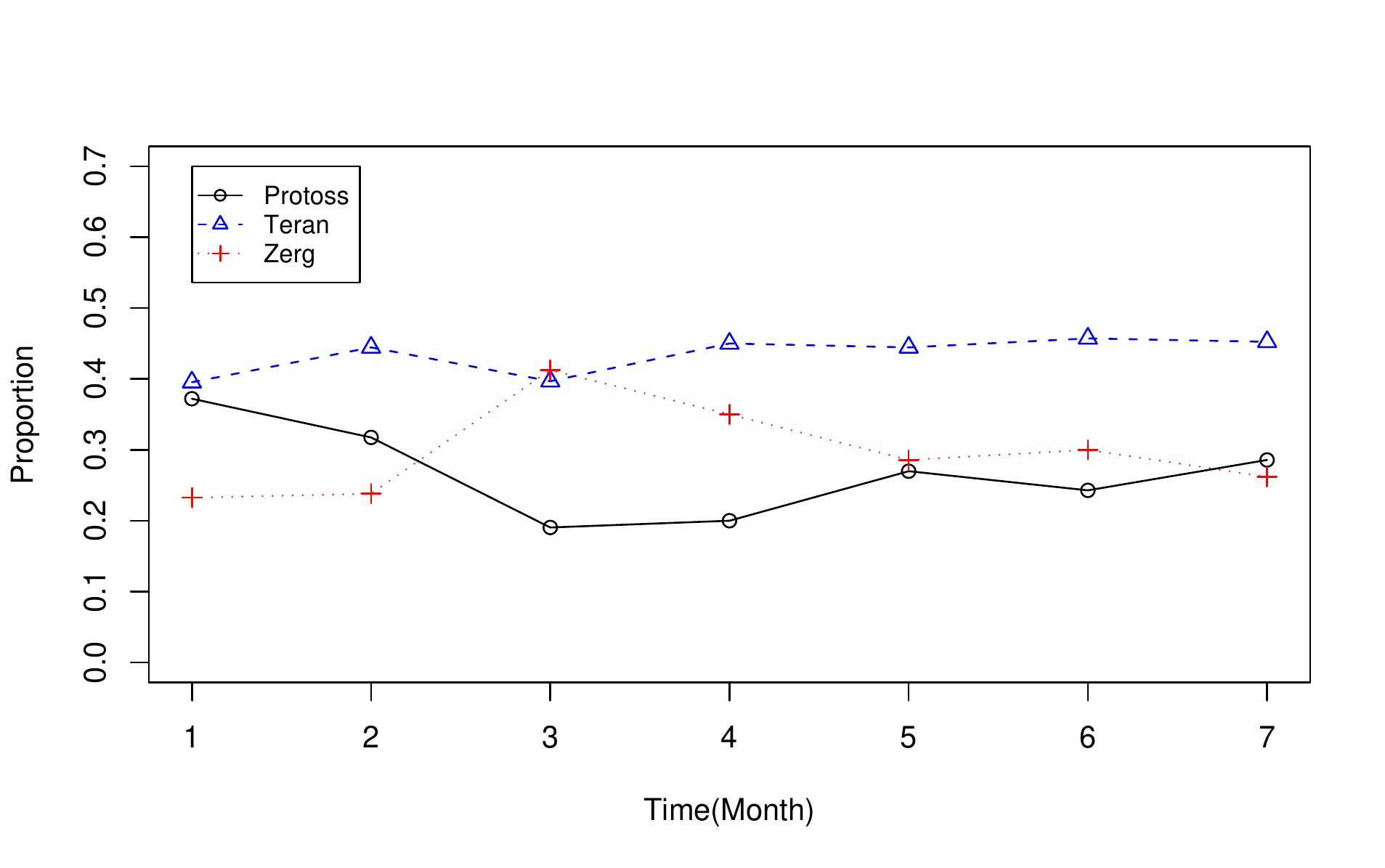}
	\caption{Proportions of races according to time}
	\label{Proportions of races according to time}
\end{figure}

Observing trend of race proportions will help understand balancing problem. We divide the data into 7 sub-data by months, and see the number of players of each race. Each cell count is frequency and the numbers in each parenthesis is the proportion of each race conditioning on each period row. Refer to the Table~\ref{Time trend of the number of players for each race} and Figure~\ref{Proportions of races according to time}.

\begin{table}[hbp!]
	\centering
	\begin{tabular}{r | r | r | r }
	\hline
	Period & Protoss Players & Teran Players & Zerg Players \\
	\hline
	September, 2010	&16 (0.37209)	&17 (0.39534)	&10 (0.23255) \\
	October, 2010		&20 (0.31746)	&28 (0.44444)	&15 (0.23809) \\
	November, 2010	&12 (0.19047)	&25 (0.39682)	&26 (0.41269) \\
	December, 2010	&4 (0.20000)	&9 (0.45000)	&7 (0.35000) \\
	January, 2011		&17 (0.26984)	&28 (0.44444)	&18 (0.28571) \\
	February, 2011	&17 (0.24285)	&32 (0.45714)	&21 (0.30000) \\
	March, 2011			&12 (0.28571)	&19 (0.45238)	&11 (0.26190) \\
	\hline
	\end{tabular}
	\caption{Time trend of the number of players for each race}
	\label{Time trend of the number of players for each race}
\end{table}

\newpage

\subsection{Appedix B: Validation check with ranks}

Table~\ref{ranks} shows the prize ranks up to March 19th, 2011. The third column is the estimated ranks based on our model.  Prize ranks more than 20 were not publicized thus not displayed.

\begin{table}[htp!]
  \centering
  \begin{tabular}{ccc}
    \hline
    Rank & Name & Rank in Prize Money (Korea Won)\\
    \hline
    1 & Min-Chul Chang & 1 \\
    2 & Yong-Hwa Choi & \\
    3 & Kang-Ho Hwang & \\
    4 & Jae-Duk Lim & 2 \\
    5 & Young-Jin Kim & \\
    6 & Jun-Sik Yang & \\
    7 & Sung-Jun Park & 8 \\
    8 & Jun-hyuk Song & 15 \\
    9 & Hyun-Woo Park & \\
    10 & Won-Ki Kim & 3 \\
    \hline
  \end{tabular}
  \caption{Parameter Estimate Rank and Prize Money Rank(up to
    March. 19. 2011).}
   \label{ranks}
\end{table}
Although it is not very coherent with the prize rank, as experts of this problem we see that this rank to be very convincing. Some of the gamers ranked high here have been recently came to the tournament, not having enough opportunities to get high prize money.

\section*{Acknowledgement}

We thank Kyungmin Ahn, Jinhak Kim, and professor
Olga Vitek for helpful comments and
contributions on descriptive analysis of data.

\end{document}